\documentclass[english,a4paper,pre,twocolumn,amsmath,amssymb,showpacs,superscriptaddress]{revtex4}

\usepackage{bmpsize}
\usepackage{graphicx,color}								
\usepackage{amssymb,amsmath}
\usepackage{natbib}
\usepackage{color}
\usepackage{array}
\usepackage{rotating}
\usepackage{subfigure}

\begin{document}

\author{Rhiannon Pinney}
\affiliation{HH Wills Physics Laboratory, Tyndall Avenue, Bristol, BS8 1TL, UK}
\affiliation{Bristol Centre for Complexity Science, University of Bristol, Bristol, BS8 1TS, UK}

\author{Tanniemola B.  Liverpool}
\affiliation{School of Mathematics, University of Bristol, Bristol, BS8 1TW, UK}
\affiliation{BrisSynBio, Tyndall Avenue, Bristol, BS8 1TQ, UK}

\author{C. Patrick Royall}
\affiliation{HH Wills Physics Laboratory, Tyndall Avenue, Bristol, BS8 1TL, UK}
\affiliation{School of Chemistry, University of Bristol, Cantock Close, Bristol, BS8 1TS, UK}
\affiliation{Centre for Nanoscience and Quantum Information, Tyndall Avenue, Bristol, BS8 1FD, UK}

\email{paddy.royall@bristol.ac.uk}

\title{Yielding of a Model Glassformer: an Interpretation with an Effective System of Icosahedra}

\begin{abstract}

We consider the yielding under simple shear of a binary Lennard-Jones glassformer whose super-Arrhenius dynamics are correlated with the formation of icosahedral structures. We recast this glassformer as an effective system of icosahedra [Pinney \emph{et al. J. Chem. Phys.} \textbf{143} 244507 (2015)]. Looking at the small-strain region of sheared simulations, we observe that shear rates affect the shear localisation behavior particularly at temperatures below the glass transition as defined with a fit to the Vogel-Fulcher-Tamman equation. At higher temperature, shear localisation starts immediately upon shearing for all shear rates. At lower temperatures, faster shear rates can result in a delayed start in shear localisation; which begins close to the yield stress. Building from a previous work which considered steady-state shear [Pinney \emph{et al. J. Chem. Phys.} \textbf{143} 244507 (2016)], we interpret the response to shear and the shear localisation in terms of a \emph{local} effective temperature with our system of icosahedra. We find that the effective temperatures of the regions undergoing shear localisation increase significantly with increasing strain (before reaching a steady state plateau).

\end{abstract}


\maketitle

\section{Introduction}
\label{sectionIntroduction}

The mechanism behind the rapid slowing in liquids approaching the glass transition has been extensively studied via a multitude of theoretical approaches, although an agreement on the nature of the liquid-to-glass transition remains elusive \cite{cavagna2009,berthier2011}. While their quiescent nature remains poorly understood, the response of a glass to mechanical stress presents further challenges  \cite{barrat2011,bonn2015}. This lack of understanding significantly hampers the exploitation of, for example, metallic glasses, an otherwise promising emergent material \cite{schuh2007,cheng2011}.

In crystalline materials, yielding is typically associated with grain boundaries -- these are defects, where materials tend to fail.  By contrast, \emph{amorphous} materials do not have readily identifiable defects, and no general theory has been established for microscopic yield mechanisms \cite{barrat2011,bonn2015,schuh2007,maass2015}. However it is understood that glasses under simple shear undergo particle displacements in two different regimes; below a certain threshold, the system lies in the \emph{transient regime} where (mostly) reversible, \emph{elastic} deformations take place. Beyond this threshold, the system undergoes irreversible, \emph{plastic} deformations to a large extent \cite{saw2016}. The threshold between these two regimes is the \emph{yield} point \cite{barrat2011,bonn2015}. Here we concern ourselves with the transient regime leading up to and beyond this yield point, characterised for our purposes by the point at which the shear stress reaches a maximum. Yielding is often accompanied by shear localisation (or shear banding), the separation of a sheared system into two regions of different viscosity \cite{barrat2011,bonn2015}.

Signatures exhibited by sheared amorphous systems in the elastic regime, such as \emph{soft modes} \cite{widmer2009,candelier2009,xu2010,mosayebi2014}, \emph{shear transformation zones}, (STZs) \cite{falk1998,falk2011,barrat2011,puosi2016}, \emph{hot spots} \cite{amon2012}, and Eshelby-like strain events \cite{eshelby1957,chattoraj2013} have been shown to play a key role in the mechanics of amorphous solids  \cite{barrat2011}. In particular, the time and location of these plastic shear deformations can be predicted using these ``soft'' and ``hard'' regions \cite{steif1982,manning2007,rottler2014,antonaglia2014}. Approaching the yield point, STZs can be power-law distributed \cite{lemaitre2009} and exhibit spatial and temporal correlations \cite{tanguy2006,besseling2010,lemaitre2009,chikkadi2012,chikkadi2014,mandal2013,chattoraj2013,benzi2014,nicolas2014} and the energy related to these plastic events can propagate through the system in an elastic fashion, leading to failure in other regions, often accompanied by shear localisation. Such behaviour is well captured by approaches such as mesoscopic elastoplastic models \cite{sollich1998,hebraud1998,martens2012,agoristas2015,lin2016} and kinetic Monte Carlo approaches \cite{homer2010}. This has also been shown in 2D Lennard-Jones systems \cite{tsamados2009,karmakar2010}.  Furthermore, shear has been used to access the so-called Gardner transition \cite{charbonneau2014} between glass states with differing stabilities \cite{berthier2016,biroli2016}.

Shearing transforms the material and has been shown to increase the potential energy of soft glassy materials \cite{bonn2002,utz2000} (subsequent relaxation can enable the system to reduce its potential energy \cite{derlet2018}), and varying the shear rate can yield systems with different \emph{effective temperatures} \cite{ono2002,berthier2002,manning2007,shi2007,manning2009,bailey2006,falk2011,maass2015,nicolas2016}. One should note that the effective temperature can be distinct from the actual temperature and that thermalisation in these systems is typically very rapid \cite{barrat2011}. In our own case, like the above, effective temperature is directly coupled to structural properties \cite{pinney2015}, so essentially what we mean by effective temperature is that the structure corresponds to the quiescent system at a temperature different to that of the simulation. We note that our approach is distinct from some previous work, in that our definition of effective temperature is based on a \emph{distribution} of structural quantities \cite{pinney2015,pinney2016}.

By analogy with crystalline solids where defects have a clear structural signature, one may infer from the above that shear localisation should be related to the structure. In the case of amorphous systems, by analogy, one 
imagines a change in local structure which would underly the shear localisation \cite{schuh2007,maass2015}. The challenge is to identify the relevant structural components. Indeed early theories of mechanical failure assume a structural origin in the form of free volume \cite{cohen1959,spaepen1977}. This is reflected in the assumption that STZs must have some particular structural characteristics which make them sucsceptible to deformation under stress \cite{falk1998,schuh2007}. It has been suggested that incorporating local structure in mesoscopic models may well improve their behaviour \cite{homer2010}. It is known that STZs can be related to local elasticity \cite{mayr2009,tsamados2009}. Here we seek a direct structural interpretation.

\begin{figure}
\begin{center}
\includegraphics[width=90mm]{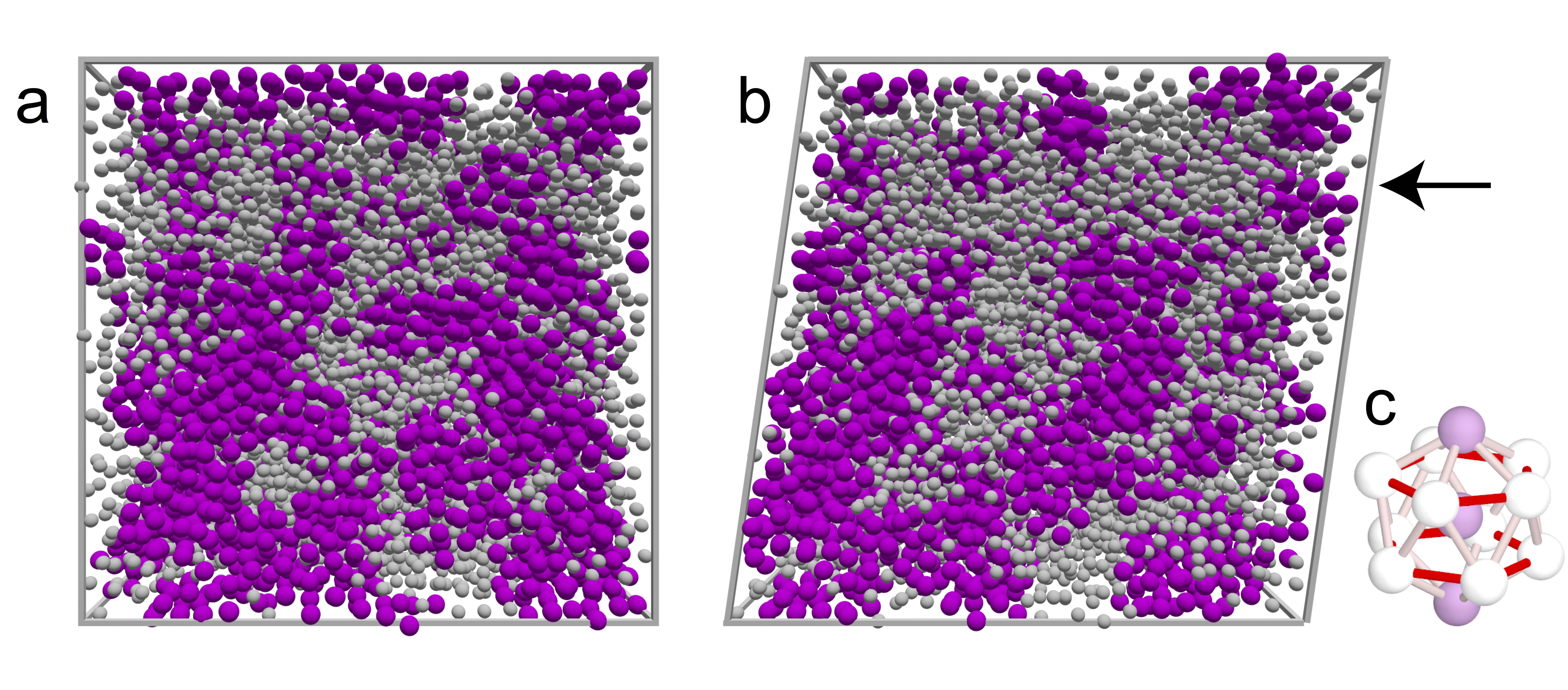}
\caption{Snapshot of the local structure in a glass at $T=0.3$, sheared at a rate of $\dot{\gamma}=10^{-5}$. Particles highlighted in magenta are in icosahedra (locally favoured structures) as depicted in (c). (a) Quiescent state prior to shear. (b) Sheared system at a strain of 0.1. Arrow denotes location of shear band, coincident with a local drop in population of icosahedra.}
\label{figPretty}
\end{center}
\end{figure}

Now in quiescent supercooled liquids, it is possible to correlate local structure, in the form of geometric motifs, such as icosahedra [Fig. \ref{figPretty}] \cite{dzugutov2002,coslovich2007,malins2013jcp,royall2017}, or more generic \emph{amorphous order} \cite{biroli2008,sausset2011,cammarota2012epl,dunleavy2012,ozawa2015} with dynamical arrest using computer simulation \cite{royall2015physrep,steinhardt1983,jonsson1988,dzugutov1992,coslovich2007,eckmann2008,lerner2009,sausset2010,tanaka2010,malins2013jcp,hocky2014,royall2015,turci2017}, particle-resolved studies in colloid experiments \cite{hunter2012,konig2005,ivlev,royall2008,mazoyer2011,leocmach2012,tamborini2015}, and high-order diffraction studies on atomic systems \cite{hirata2009,liu2013}. In particular, such locally favoured structures, (LFS) have been correlated with the slow regions of the dynamic heterogeneity exhibited by glassy systems \cite{dzugutov2002,coslovich2007,tanaka2010,sausset2010,malins2013jcp,jack2014,hocky2014,turci2017}. although this correlation alone does not constitute a mechanism for arrest \cite{charbonneau2013,jack2014}.

Given that the emergence of solidity can now be related to local structure in the amorphous state, one is motivated to enquire as to whether sheared systems might exhibit similar behaviour, with fewer geometric motifs in the ``soft spots''. The role of the initial structural state, in controlling the mechanical response was recognized some time ago \cite{albano2005,shi2005,shi2007} and has since been related to features such as soft spots \cite{ding2014,schoenholz2014} and shear bands \cite{ding2012,feng2015}. Much remains to be done, as in experiments on metallic glass, identifying shear bands can be challenging in bulk specimens, as it is very hard to identify the microstructural changes that occur following the shear localisation \cite{schuh2007,hassani2016}.

Now direct detection of LFS requires particle-resolved colloidal experiments or computer simulations, which are restricted to the first 4-5 decades of dynamic slowing, compared to 14 decades required to reach the glass transition ($T_{g}$) in atomic and molecular systems \cite{berthier2011,royall2015physrep}. Understanding the glass transition therefore necessitates data extrapolation far beyond the accessible regime \cite{debenedetti2001}. We emphasise that $T_g$ is distinct from lower temperatures at which the relaxation time of the material may diverge according to some interpretations \cite{berthier2011}, such as that predicted by Adam-Gibbs \cite{adam1965} and Random-First-Order Transiiton theories \cite{lubchenko2007} and captured by the Vogel-Fulcher-Tamman (VFT) expression, $T_{\mathrm{VFT}}$ \cite{berthier2011,royall2015}.

In a previous publication, we addressed the challenge presented by the limited timescales available to computer simulation and recast a well-studied binary Lennard-Jones glassformer into an effective system of LFS to develop a population dynamics model of domains of icosahedra which we term \emph{mesoclusters} \cite{pinney2015}. The ideas behind our model are illustrated in Fig. \ref{figPretty}, where particles identified in icosahedra or otherwise are rendered accordingly. In the quiescent system [Fig. \ref{figPretty}(a)], below a certain temperature, (around 0.62 in the system of interest \cite{malins2013jcp}) the icosahedra form a percolating network. Our model predicts the size distribution of the mesoclusters of icosahedra, incorporating the effects of percolation \cite{pinney2015}. Coupling the lifetime of mesoclusters of a certain size with this size distribution, the model successfully describes the increase in relaxation time upon cooling and can be used to predict system behavior at significantly colder temperatures than those accessible to our simulations. By construction, our model does not predict a thermodynamic phase transition to an ``ideal glass''. Under steady-state shear, (\emph{i.e.} beyond the yield point, well into the plastic regime) using our mesocluster size model \cite{pinney2015}, we were able to obtain effective temperatures for the system due to the change in the distribution of of mesoclusters of icosahedra \cite{pinney2016}. We were thus able to link local structure to the local effective temperatures observed previously \cite{ono2002,berthier2002,manning2007,manning2009,bailey2006,falk2011,maass2015,nicolas2016}, with a key difference, for in our case, determining effective temperature means mapping a \emph{distribution} of mesocluster sizes, rather than a single parameter.

Here we turn our attention to the \emph{elastic} regime, for small strains below yielding. In particular we consider the proposition that a sheared system exhibits mesocluster size distributions that are well described by our model (with \emph{no} changes to the parameterization).  In other words, we make the assumption that the sheared system may be treated as a quiescent system  with a different (effective) temperature. Clearly such simplification neglects effects such as local dilation due to shearing, and other specific shear-induced structural, changes, such as distortion of LFS \cite{albano2005}.

We consider the Wahnstr\"{o}m binary Lennard-Jones glassformer \cite{wahnstrom1991} under an imposed uniform planar shear for total strain $\leq 30\%$. We aim to understand sheared systems with simulation temperatures both above and below $T_{\mathrm{VFT}}$ with an \emph{effective} temperature fitted using our mesocluster size model \cite{pinney2015}. The obtained effective temperatures are tracked with strain as the system is sheared and we find that the effective temperature rises with increasing strain. Interestingly, most of the state points which exhibit shear localisation do so immediately upon the initialisation of shear, with a notable exception being very cold state points ($T \leq 0.3$) where the localisation begins at approximately the yield strain which is distinct from the behaviour in a true glass in \emph{experimental} timescales \cite{schuh2004}, while here we have a supercooled liquid-glass crossover. We also show that the localisation behavior of simulation runs with the same initial particle configuration are significantly different for varying shear rates. 
We find that regions undergoing shear localisation have distinctly higher local effective temperatures relative to the rest of the system.

Our strategy here is to focus on obtaining configurations as deeply supercooled as possible. Certain theories of the glass transition, such as the Random First Order Transition theory,  emphasise a change in the nature of the relaxation dynamics upon deep supercooling (past the so-called Mode-Coupling crossover) \cite{lubchenko2007,wisitsorasak2017}. We therefore have elected to focus our computational resources upon obtaining systems equilibrated at as low a temperature as possible, in particular at temperatures lower than the Mode-Coupling Crossover, $T_\mathrm{mct}\approx0.57$ for our system \cite{lacevic2003,pinney2015}. Thus we choose to use rather modest (by some measures) system sizes of $N=10,976$ particles. In particular, we equilibrate the system to a temperature of $T=0.56$ (past the Mode-Coupling Crossover), with the aim of accessing a more deeply supercooled regime.

This paper is organized as follows: in our methododology section (section \ref{sectionMethodology}), we briefly recap our mesocluster model in (section \ref{sectionRecap}) and describe the simulation protocol and analysis method in (section \ref{sectionSimulation}). We divide the results (section \ref{sectionResults}) into four. Section \ref{sectionTransient} shows our identification of the yield stress and yield strain for our simulations. We discuss our methods for detecting local shear in section \ref{sectionDetermining} and how we use these to identify shear bands in section \ref{sectionCriteria}. We show the effective temperature under shear localisation in section \ref{sectionEffective} and discuss and interpret our results in Section \ref{sectionDiscussion}. We conclude with a summary in Section \ref{sectionSummary}.

\begin{figure*}[!htb]
\begin{center}
\includegraphics[width=180 mm]{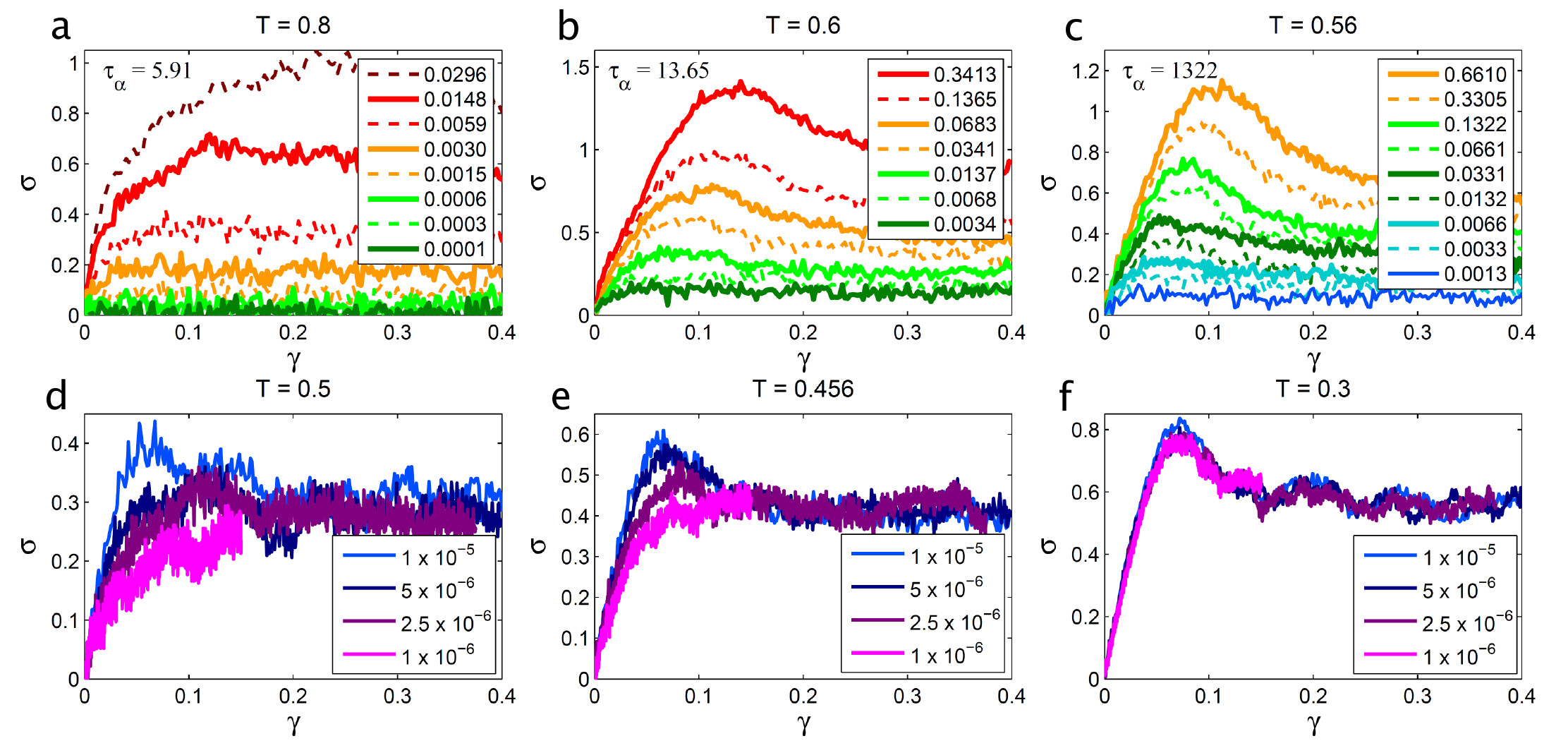}
\caption{Stress values with increasing strain for state points with $0.3 \leq T \leq 0.8$ and varying shear rate. 
(a-c) Supercooled liquids. Shear rate given in terms of the $\alpha$-relaxation time multiplied by the shear rate, $\tau_{\alpha} \dot{\gamma}$ in the legend. 
(d-f) Glass. Shear rate given in simulation units. 
Line colours and types represent equivalent shear rates in simulation units (note that the slowest shear rate in the plot of (c) $T = 0.56$ corresponds to the fastest shear rate in (d-f)). Fast shear rates have measurable transient regions, \emph{i.e.} exhibit a maximum/plateau stress value, $\sigma$, at non-zero strain, $\gamma$. Slow shear rates for temperatures $T \geq 0.56$ reach a steady stress value almost instantly. A clear yield point is visible for all shear rates in $T = 0.3$ and faster shear rates at $T = 0.456, 0.5$, but there is no clear yield point for the slowest shear rate ($10^{-6}$) in $T = 0.456, 0.5$.
Data are averaged over 8 simulations in each case.}
\label{figBigStressPlot}
\end{center}
\end{figure*}

\section{Methodology}
\label{sectionMethodology}

\subsection{Recap of population dynamics model}
\label{sectionRecap}

First, we briefly introduce the population dynamics model which generates the mesocluster size distribution from our previous work. \cite{pinney2015}. Mesoclusters are structures made up of particles in icosahedra, the LFS for the Wahnstr\"{o}m model glassformer \cite{coslovich2007,malins2013jcp}. We assume that mesoclusters of size $m$ ($m$ being the number of icosahedra, expressed as the number of particle at the centre of an icosahedron) can only change in size by $\pm 1$ and are restricted in size by a system-size dependent constant $M$. For high temperatures, $p_{m}$ (the probability of a mesocluster being size $m$) follows an exponential decay with steady-state solution 
\begin{equation}
p_{m}(T)= a(T)^{m-1}p_{1}(T) 
\end{equation}
where $a(T)$ is the temperature-dependent decay parameter. At lower temperatures below $T=0.62$ \cite{malins2013jcp,pinney2015}, the mesoclusters percolate Fig. \ref{figPretty}(a), and as such the shape of their size distribution changes. We account for this change by including a Gaussian weighting to obtain the steady state solution 
\begin{equation}
p_{m}(T) = a(T)W_m(T)p_{m-1}(T) 
\label{eqSteadyState}
\end{equation}
where $a(T)$ is an underlying decay parameter and $W_{m}(T)$ is the Gaussian weight which include ``mean'' and ``variance'' parameters to control the shape of the distribution.

One can then use the population of mesoclusters of icosahedra to predict the dynamics of the quiescent system. In particular, we make the assumption that the super-Arrhenius increase of the structural relaxation time $\tau_\alpha$ is given by the population of of mesoclusters of icosahedra and the lifetime of each size of mesocluster $l_m$:

\begin{equation}
\tau_\alpha = \tau_\alpha^\mathrm{Arr} \sum_{m}l_m(T)p_m(T)
\label{eqFavourite}
\end{equation}

\noindent were $\tau^\mathrm{Arr}$ is the relaxation time assuming Arrhenius behaviour, extrapolated from the high-temperature $T>1$ behaviour. Each icosahedron is categorised according to the largest mesocluster it joins during its lifetime. The lifetime of an icosahedron is determined by the amount of (simulation) time that has elapsed between the first and last instance of an icosahedron being identified by the topological cluster classification \cite{malins2013tcc}. Remarkably, we found that this simple expression in Eq. \ref{eqFavourite} gave an accurate prediction of the structural relaxation time through the regime accessible to our computer simulations \cite{pinney2015}.

\vspace{10mm}
\subsection{Simulation and Analysis} 
\label{sectionSimulation}

We simulate the Wahnstr\"{o}m equimolar additive binary Lennard-Jones model \cite{wahnstrom1991}. The size ratio is $5/6$ and the well depth between all species is identical. The mass of the large particles is twice that of the small. We use molecular dynamics simulations of $N=10976$ particles. We equilibrate for at least $100 \tau_\alpha$ in the NVT ensemble for $0.56 \leq T \leq 0.8$ and use the final configuration for $T = 0.56$ to initiate further NVT simulations at temperatures $0.3 \leq T \leq 0.5$ for as long as computationally possible. Here $\tau_\alpha$ is the structural relaxation time, determined from a fit to the intermediate scattering function \cite{pinney2015}. Throughout we work in reduced units, following \cite{malins2013jcp}.

The final configuration of each simulated temperature is used as the initial configuration of a sheared simulation following the SLLOD algorithm with Lees-Edwards periodic boundary conditions. A total of 8 simulations were produced from different configurations of each state point. All of these sheared simulations were carried out using LAMMPS \cite{LAMMPS}. The shear rates studied (in simulation units) are: $1 \times 10^{-5} \leq \dot{\gamma} \leq 0.25$ for $0.56 \leq T \leq 0.8$ and $2.5 \times 10^{-6}$, $5 \times 10^{-6}$ and $1 \times 10^{-5}$ for $0.3 \leq T \leq 0.5$. We restrict our simulations to total strain $\leq 40\%$.

We identify icosahedra with the topological cluster classification (TCC) and consider those which last longer than $0.1\tau_\alpha$ (for $0.56 \leq T \leq 0.8$) or longer than 150 simulation time units (for $0.3 \leq T \leq 0.5$) to suppress the effects of thermal fluctuations. Our structural analysis protocol is detailed in Ref. \cite{malins2013tcc}. We express the local structure as the proportion of particles identified in icosahedra, $\phi$.

\section{Results}
\label{sectionResults}

\begin{figure*}[!htb]
\begin{center}
\includegraphics[width=180 mm]{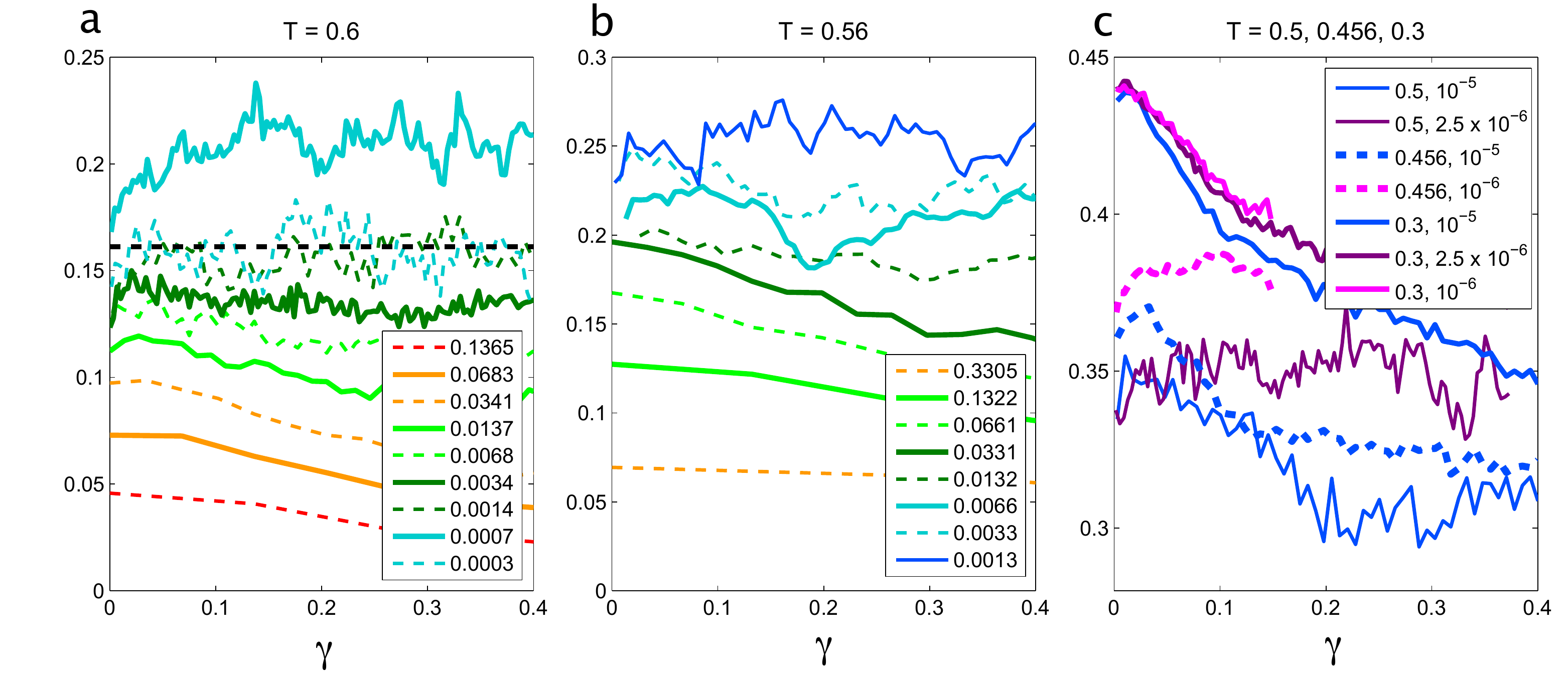}
\caption{Populations of icosahedra under increasing strain for state points with $0.3 \leq T \leq 0.8$ and varying shear rate. 
Line colours and types represent equivalent shear rates in simulation units (note that the slowest shear rate in the plot of (c) $T = 0.56$ corresponds to the fastest shear rate in (d-f) Each line is averaged over 8 realisations of each state point except for $T=0.50$.} 
\label{figNewTransientPhi}
\end{center}
\end{figure*}

\subsection{Global behaviour in the Transient Regime}
\label{sectionTransient}

We begin our discussion of the results by considering the overall response of the system. The period between initialisation of shear from a quiescent state up to the yield point is called the \emph{transient} regime. We identify this region in our simulations by looking at where each of the stress-strain curves reach a maximum value/plateau. The maximum value and its corresponding strain value is determined by fitting the peak to a quadratic function. The existence of the stress peak could be explained using arguments based on \emph{free volume} (\emph{i.e.} the amount of space available for particles to move around). If the free volume at initialisation is less than that of the steady-state, the stress must first increase to a value higher than the steady stress. This increased stress is needed to initiate a shear transformation, which may increase the free volume, which in turn aids further shear transformations and a reduction in the stress until it reaches its steady state \cite{jiang2015}. The value of this maximum stress is determined by temperature and shear rate (faster shear rate, larger maximum value) \cite{koumakis2012} consistent with previous studies of 3D Lennard-Jones systems under simple shear \cite{rottler2003,varnik2004}. Figure \ref{figBigStressPlot} shows these curves for $0.3 \leq T \leq 0.8$.

In simulations at $T = 0.8$ [Fig. \ref{figBigStressPlot}(a)], the stress-strain plots indicate that the system acts much more like a liquid, with the stress reaching a plateau very quickly after shear initialisation. Note that here the strain rate is scaled by the $\alpha-$relaxation time and here $\dot{\gamma} \tau_\alpha \ll 1$ so the liquid-like behavior is reasonable. At a temperature $T = 0.6$  [Fig. \ref{figBigStressPlot}(b)],  we see stress peaks (yield points) for the faster shear rates. These disappear with decreasing shear rate, meaning that the system yields immediately and does not have an observable transient region. At $T = 0.56$  [Fig. \ref{figBigStressPlot}(a)], the stress-strain plots have clearly identifiable yield points, and thus observable transient regions for all except the slowest shear rates studied. The yield strain patterns observed here are consistent with those found in other simulations \cite{rottler2003,varnik2004} and experiments \cite{sentjabrskaja2013,lu2003} of glassy materials.

The bottom row of Fig. \ref{figBigStressPlot} shows the stress-strain curves for $T = 0.5, 0.456, 0.3$ (note that for this system, the temperature at which the structural relaxation time is predicted to diverge according to the Vogel-Fulcher-Tamman equation for the relaxation time is $T_{\mathrm{VFT}} = 0.456$). 
\begin{equation}
\tau_\alpha(T)=\tau_0 \exp \left(  \frac{D}{T-T_{\mathrm{VFT}}} \right)
\label{eqVFT}
\end{equation}

\noindent 
where $\tau_0$ is a microscopic timescale and $D=0.799$ \cite{pinney2015} is the \emph{fragility parameter} \cite{royall2015physrep,berthier2011}. Here we take  $T_{\mathrm{VFT}}$ as the glass transition. Thus the data are plotted as $\dot{\gamma}$ rather than $\dot{\gamma} \tau_\alpha$. The yield points are clearly visible in all $T = 0.3$ data, but are less clear in $T = 0.5$ and the slower shear rates in the case of $T = 0.456$. In particular, the slowest shear rates of $T = 0.5$ and $T = 0.456$ do not have clearly identifiable yield points, and the stress increases as the strain passes beyond what would have been the yield point at faster shear rates. The yield strain for all temperatures $ \geq 0.456$ decreases with shear rate, and the yield stress shows a clear temperature and shear rate dependence; the yield stress increases with decreasing temperature and/or increasing shear rate.

\begin{figure*}
\begin{center}
\includegraphics[width=14cm]{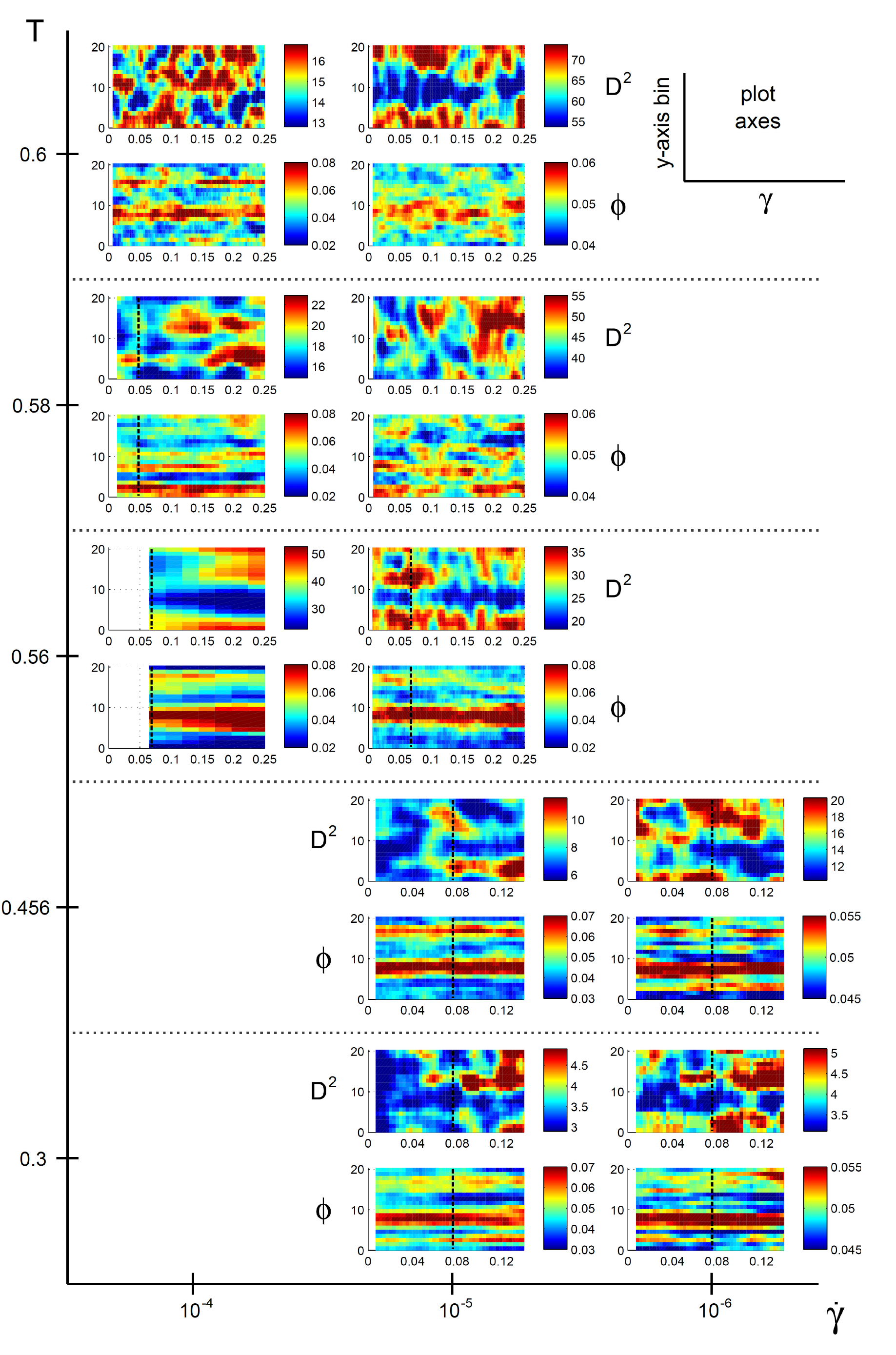}
\caption{Non-affine displacement $D^2$ values and icosahedra densities for small strain of state points with shear rates (in simulation units) $\dot{\gamma} = 10^{-4}, 10^{-5}$ for temperatures $T = 0.6, 0.58, 0.56$ and $\dot{\gamma} = 10^{-5}, 10^{-6}$ for temperatures $T = 0.456, 0.3$. Data are examples from a single simulation of each state point. Yield points (where calculable) are indicated by dashed black lines. State points are separated by a dotted grid. For each state point: (top) $D^2$ profile for small strain region of simulation and (bottom) normalised icosahedra density.}
\label{figTransientD2BigFig}
\end{center}
\end{figure*}

\subsection{Determining Local Shear}
\label{sectionDetermining}

We now turn our attention to the local behaviour, with a particular focus on identifying shear localisation. As shown and studied in Ref. \cite{pinney2016}, the low temperature state points ($T \leq 0.6$) exhibit shear localisation in the steady state. The local shear rate of each band is strongly correlated with the local icosahedra population, \emph{i.e.} many icosahedra (and/or large mesoclusters) are found in regions with low shear rate, and few icosahedra (and/or small mesoclusters) are found in regions with high shear rate. By high and low shear rates, we are making a relative comparison of the local degree of shear. A region undergoing shear localisation has a high shear rate and one not has a low shear rate. In Ref. \cite{pinney2016}, we compared $D^2_\mathrm{min}$ with the local shear rate and found the former to be a better measure. Here we continue to use $D^2_\mathrm{min}$.  We use the non-affine displacement parameter, $D^2_\mathrm{min}$ (defined in Eq. \ref{eqD2}), \cite{falk1998,gannepalli2001}, to define a localisation criterion to define \emph{when} the shear bands start. For ease of later notation, we henceforth refer to $D^2_\mathrm{min}$ as simply $D^2$.

\begin{equation}
D^{2}(\tau, t) = \sum_{n=1}^{N} \textbf{R}_{n} \cdot \textbf{R}_{n}^{T}
\label{eqD2}
\end{equation}

\begin{align}
\textbf{R}_{n} &= \Big(\textbf{r}_{n}(t) - \textbf{r}_{0}(t) \Big) - \Big(\textbf{XY}^{-1} \Big) \cdot \Big(\textbf{r}_{n}(\tau) - \textbf{r}_{0}(\tau) \Big) \nonumber \\
\textbf{X} &= \sum_{n=1}^{N} \Big(\textbf{r}_{n}(t) - \textbf{r}_{0}(t) \Big) \Big(\textbf{r}_{n}(\tau) - \textbf{r}_{0}(\tau) \Big) \nonumber \\
\textbf{Y} &= \sum_{n=1}^{N} \Big(\textbf{r}_{n}(\tau) - \textbf{r}_{0}(\tau) \Big) \Big(\textbf{r}_{n}(\tau) - \textbf{r}_{0}(\tau) \Big)
\label{eq:D2extras}
\end{align}

The shear band location was defined as follows \cite{pinney2016}: The simulation box was segmented along the $y$-axis to form 20 equal bins of roughly 1 particle diameter in height. Each bin is characterized by the average $D^2$ value of all the particles residing within that bin. To quantify whether or not the system is undergoing shear localisation, we compare the average range of different $D^2$ values observed across the $y$-axis with the average range of $D^2$ values observed within each $y$-axis slice through time:

\begin{equation}
R = \frac{\left\langle D^{2}_{\mathrm{max}} - D^{2}_{\mathrm{min}} \right\rangle_{y}}{ \left\langle D^{2}_{\mathrm{max}} - D^{2}_{\mathrm{min}} \right\rangle_{t}}
\label{eqbandingcriteria}
\end{equation}

\noindent where subscripts $y,t$ are the parameters to be averaged over; $y$-axis and time respectively. The value of $R$ quantifies how strongly banded the system is. Strong localisation is characterized by large values of $R$; where the variation in $D^2$ values along the $y$-axis (significantly) exceeds the variation observed through time along that $y$-axis location. State points where the system appears to fluctuate between exhibiting shear localisation and not exhibiting localisation through time have values $0.9 \lesssim R \lesssim 1.1$. $R < 0.9$ suggests that the system is not undergoing localisation at all; \emph{i.e.} the variation through time at a particular $y$-axis location exceeds that of the variation observed across the $y$-axis.

Figure \ref{figTransientD2BigFig} shows examples of the $D^2$ values and corresponding icosahedra populations for state points with $0.3 \leq T \leq 0.6$ for different shear rates but with the same initial configuration used for each temperature. Shown in Fig. \ref{figTransientD2BigFig} are values along the $y$-axis over which the system is sheared. From the $D^2$ profiles, we can see that those state points which exhibit shear localisation do so instantaneously with the initialisation of shear. In addition, the high/low shear regions correspond to the regions of low/high icosahedra population.

For $T = 0.6$, the system exhibits very different $D^2$ profiles when the shear rate is varied. This is despite the icosahedra density profile being (qualitatively) similar. The behaviour for $T = 0.58, 0.56$ look similar across both shear rates, with $T = 0.58$ showing banded regions which change with time and $T = 0.56$ having shear localisation more constant in time that are strongly correlated with the icosahedra density profile.

The $D^2$ data for $T = 0.456$ is strikingly different for the two shear rates shown. For the faster shear rate ($10^{-5}$), we see a clear increase in the maximum $D^2$ value close to the yield point which persists for the remaining time shown. Note that for these low temperatures, $T\leq T_\mathrm{VFT}$ the structural relaxation time is not defined, and we quote the shear rate in simulation units (see section \ref{sectionSimulation}. For the slower shear rate ($10^{-6}$), the system separates into high/low $D^2$ values instantly. In addition to this, the density of icosahedra for the slower shear rate is more homogeneous than its faster counterpart, although qualitatively the profiles are similar (recalling that the same initial configuration is used for both). Temperature $T = 0.3$ shows a similar behavior to $T = 0.456$, but the change in $D^2$ maximum values at the yield point in the faster shear rate is more pronounced.

\subsection{Criteria for Shear Localisation}
\label{sectionCriteria}

It is clear from the $D^2$ values shown in Fig. \ref{figTransientD2BigFig} that the system undergoes shear localisation. We now consider how to quantify the localisation. Using our $D^2$ values and the localisation criteria outlined in the preceding section and detailed in Ref. \cite{pinney2016}, we can identify when localisation occurs in each simulation run. Since fluctuations in $D^2$ along the $y$ axis are present in all runs (including those that do not exhibit localisation), we need a second criterion to define \emph{when} the $D^2$ values are sufficient to identify onset of shear localisation.

Using values obtained from quiescent data, we can find the ``baseline'' fluctuations of $D^2$ expected at different temperatures. We achieve this by calculating $D^{2}_{\mathrm{max}} - D^{2}_{\mathrm{min}}$ across the $y$ axis for each timestep, and normalising by the average $D^2$ value in that timestep:

\begin{equation}
D^{2}_{\mathrm{Range}} = \frac{D^{2}_{\mathrm{max}} - D^{2}_{\mathrm{min}}}{\left\langle D^2 \right\rangle}
\end{equation}

\begin{figure}
\begin{center}
\includegraphics[width=\columnwidth]{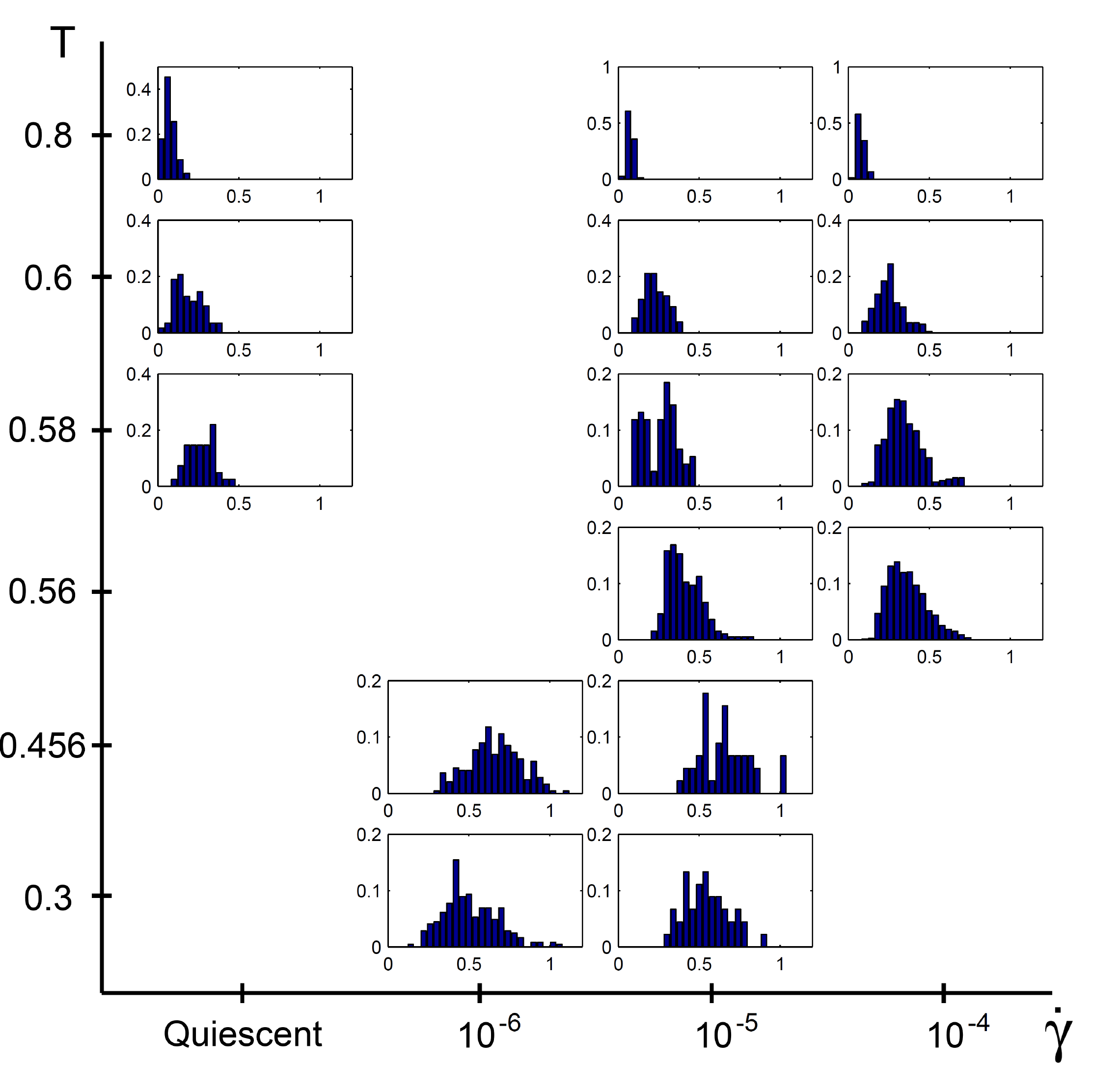}
\caption{Distributions of the (normalised) range of $D^2$ values observed across the $y$ axis in sheared (middle and right) and quiescent (far left) at varying temperatures. There is a marginal difference in the $D^2$ ranges at temperatures $T \leq 0.6$; the ranges in sheared systems are slightly larger.}
\label{figD2RangesBigFig}
\end{center}
\end{figure}

This yields a proportional change of $D^2$ from its average, \emph{i.e.} a value of 0.25 means that $D^{2}_{\mathrm{max}} - D^{2}_{\mathrm{min}}$ is $25\%$ of $\left\langle D^{2} \right\rangle$. Examples of the distributions of the range of values $D^2$ takes across the $y$ axis in some sheared and quiescent simulation runs \emph{are} 
shown in Fig. \ref{figD2RangesBigFig}. For quiescent state points the average (normalised) ranges of $D^2$ are 0.109, 0.226 and 0.305 for $T = 0.8, 0.6$ and $0.58$ respectively; the range of $D^2$ increases as temperature is decreased.

Looking at the sheared systems, Fig. \ref{figD2RangesBigFig} shows the distributions of the (normalised) range of $D^2$ values in state points with $0.3 \leq T \leq 0.8$ and shear rates (in simulation units) $\dot{\gamma} = 10^{-6}, 10^{-5}, 10^{-4}$. The $D^2$ ranges are marginally larger in sheared than in quiescent systems, but change very little with increasing the rate of shear. These sheared systems also show that the range of $D^2$ increases with decreasing temperature, at least until $T = 0.456$, where the distributions of $T = 0.456$ and $T = 0.3$ look very similar.\\

\begin{figure}
\begin{center}
\includegraphics[width=\columnwidth]{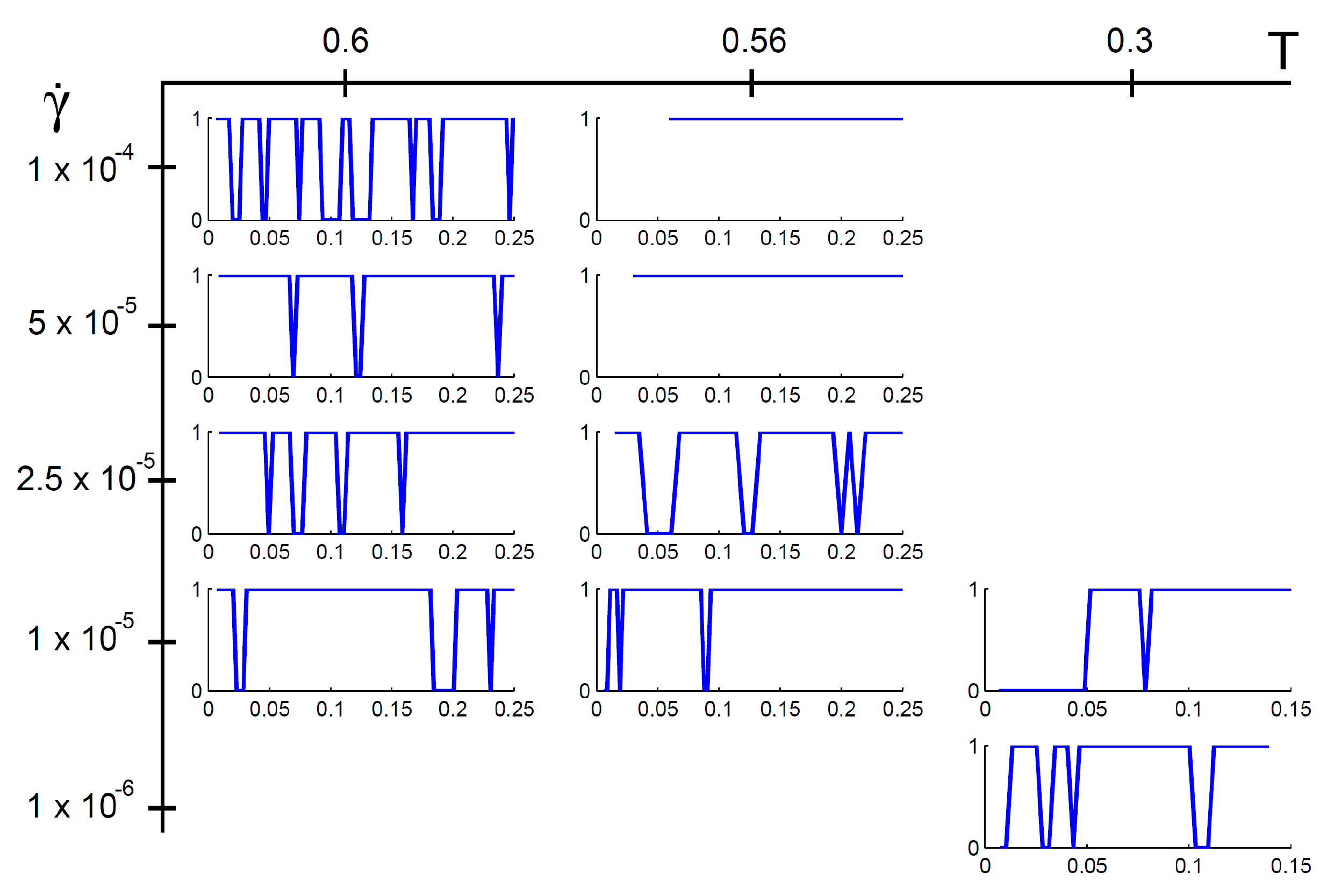}
\caption{An on/off switch indicating the prescence of localisation in for state points $T = 0.6, 0.56, 0.3$ and varying shear rates (given in simulation units). Data are from a single simulation of each state point. Values of one mean that the localisation criteria has been satisfied (\emph{i.e.} $D^2$ range $> 25\%, 35\%, 40\%$ for $T = 0.6, 0.56, 0.3$ respectively). Values of zero mean that this localisation criteria has not been satisfied.}
\label{figBandingStartAllTemps}
\end{center}
\end{figure}

\begin{figure}
\begin{center}
\includegraphics[width=
\columnwidth
]{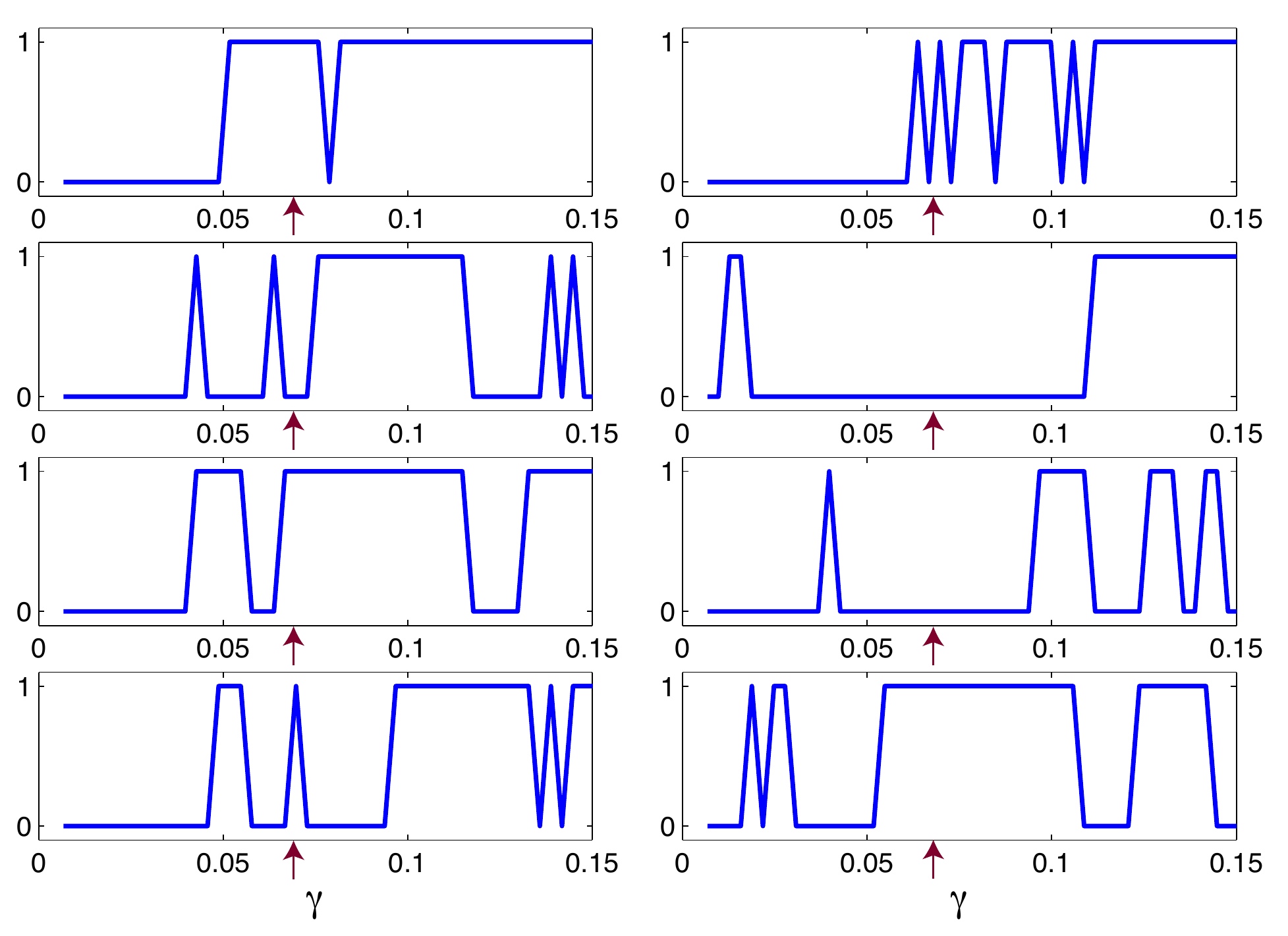}
\caption{An on/off switch indicating the presence of localisation for simulation runs with $T = 0.3$ and shear rate $\dot{\gamma} = 10^{-5}$ for all 8 different initial configurations. We can clearly see the start of near continuous shear localisation at small strain values ($\approx 0.07$) after a period of no localisation as indicated by the arrows.}
\label{figBandingStart}
\end{center}
\end{figure}

Using this data, we define the criteria for the start/end of a period of shear localisation as the point at which the $D^2$ ranges rise/fall above/below some threshold. If the sheared simulation has a $D^2$ range which exceeds that of the quiescent system, it must surely be shear localisation. Conversely, if the range is smaller, then the $D^2$ values across the simulation box are more homogenous than the quiescent system, which suggests that the system is shearing more uniformly (thus contradicting localisation). Given that the quiescent data for $T = 0.6, 0.58$ has a ``typical'' range of $25\%$ and $30\%$ (respectively) of the average $D^2$ value (and the range increases with decreasing $T$), we set the localisation threshold at $25\%$ for $T = 0.6$, $30\%$ for $T = 0.58$, $35\%$ for $T \leq 0.56$ and $40\%$ for $T \leq 0.456$. These thresholds define where the system undergoes shear localisation. Choosing higher threshold values creates a stricter criterion, while lowering them relaxes it. Selecting a higher threshold value would, in some cases (for example $T = 0.6$), result in less localisation being observed. However, in the temperature regime of interest where localisation is exhibited for the majority of the simulation time ($T \leq 0.56$), the threshold is exceeded by a significant margin. In these cases, only a particularly strict criterion would affect the results. Since we are in the transient region, the time averages of $D^2$ are ill-defined, so it is hard to evaluate the long-time $D^2$ averages used to determine the existence of localisation in Ref. \cite{pinney2016}. We will therefore use only this $D^2$ range threshold criteria within these transient region simulations. We implement this strategy in Figs. \ref{figBandingStartAllTemps} and \ref{figBandingStart}.

Figure \ref{figBandingStartAllTemps} shows the existence of banding at $T = 0.6, 0.56, 0.3$ for different shear rates. For all temperatures $T \geq 0.456$, the shear bands generally start immediately or very shortly after shear initialisation. For $T = 0.3$, $\dot{\gamma} = 10^{-5}$ there is no shear localisation until (approximately) the yield strain is reached.

Figure \ref{figBandingStart} shows the start of shear localisation for simulation runs with $T = 0.3$, $\dot{\gamma} = 10^{-5}$ for all 8 different initial particle configurations. These plots show that, at least for the state point $T = 0.3, \dot{\gamma} = 10^{-5}$, sustained shear localisation begins close to the yield point.

In both cases where shear localisation is exhibited immediately or delayed, shear localisation is very strongly correlated with the icosahedra poor regions. This provides good evidence for a link between local LFS population and local shear response. In the case of shearing glasses and deeply supercooled liquids, we see in Fig. \ref{figTransientD2BigFig} although shear does change the structure prior to yielding, the effect of the initial configuration is still significant. Note that we use deterministic molecular dynamics. Were we to carry out these kinds of shear with a Brownian dynamics system for example, the link to the initial structure might be weaker. We also see that following yielding, shear localisation regions remain up to a strain of 0.25 at least.

\subsection{Effective Temperatures}
\label{sectionEffective}

\begin{figure}
\begin{center}
\includegraphics[width=65mm]{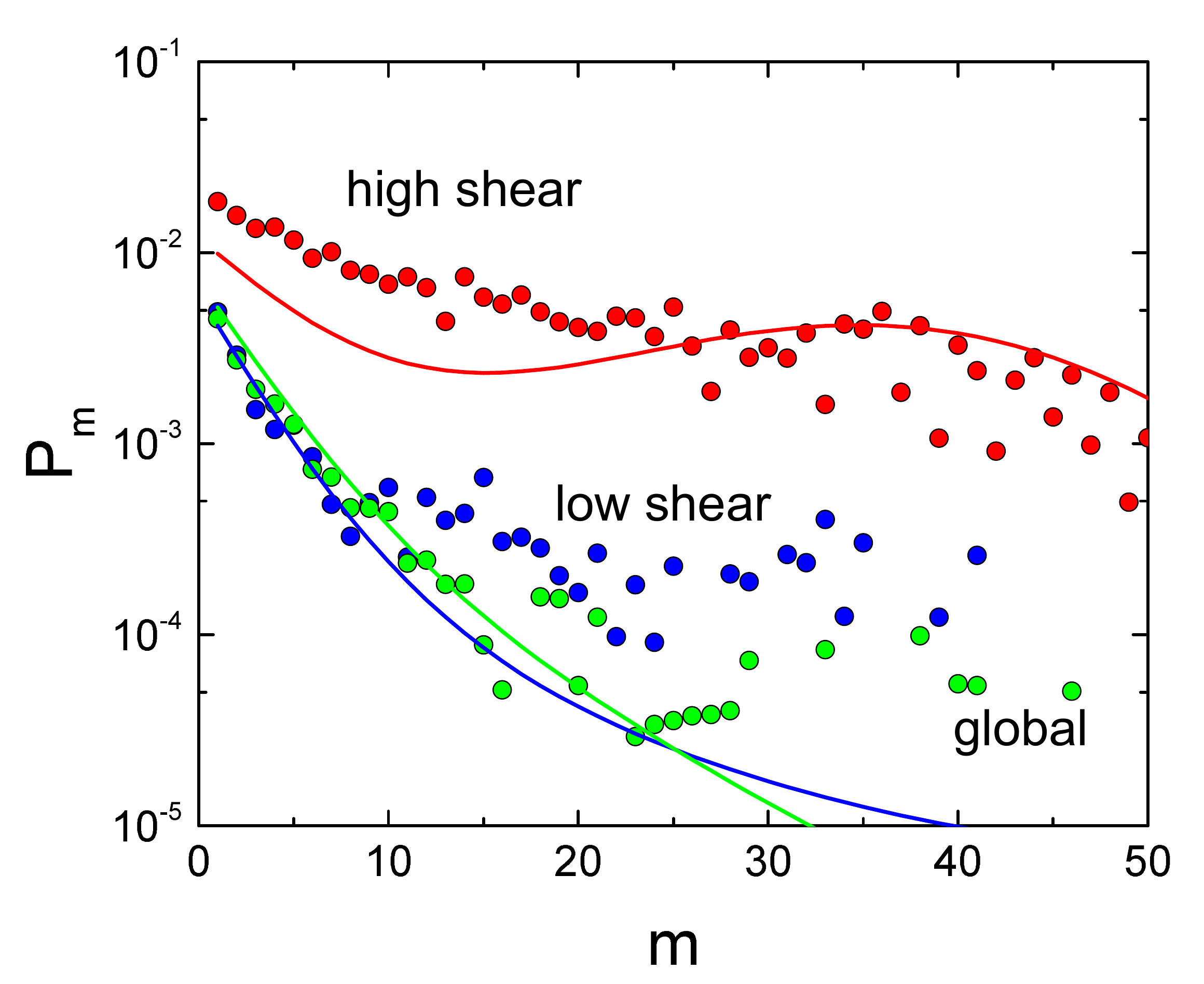}
\caption{Fitting the mesocluster size distribution (lines) to simulations (data points) obtain an effective temperature. Shown is a global fit (green) and fits to high (red) and low (blue) shear regions. The state point was $T=0.3, \dot{\gamma} = 10^{-5}$.}
\label{figEffective}
\end{center}
\end{figure}

Our model \cite{pinney2015} allows us to predict the population, $\phi$, and size distribution of mesoclusters formed of particles in icosahedra for a given temperature. Here we use the model in a reverse fashion, to determine an \emph{effective} temperature in the sheared case but taking the mesocluster size distribution and population as inputs. Thus, using the same approach as Refs. \cite{pinney2015,pinney2016}, where possible, we assign an effective temperature to the whole simulation box (when there is no localisation) and to the high/low shear segments (when the system exhibits shear localisation) using the mesocluster size model. We recall that, given the structural nature of the model, what we mean by effective temperature, is that the higher order structure, locally, through the distribution of mesoclusters of icosahedra corresponds to that in the quiescent system at some different temperature. Example fits are shown in Fig. \ref{figEffective}.

Figure \ref{figNewTransientPhi} shows the proportion of particles in icosahedra, $\phi$ in sheared simulations with $T = 0.56, 0.456$, and $0.3$. For $T = 0.56$, the icosahedra population $\phi$ steadily increases with decreasing shear rate to meet some upper limit. The values of $\phi$ observed in $T = 0.456, 0.3$ have noticeably different behaviour. Looking at $T = 0.3$, both shear rates follow roughly the same pattern of increasing $\phi$ a small amount, peaking at strain $\gamma \approx 0.01$, then falling at larger strain. This pattern appears exaggerated in $T = 0.456$ where $\phi$ increases by a larger amount and peaks at larger strain values (significantly larger for shear rate $\dot{\gamma} = 10^{-6}$) before eventually falling again.

Using the localisation threshold criterion ($D^2$ range exceeding $25\%, 30\%, 35\%$ and $40\%$ for temperature $T = 0.6, 0.58, 0.56$ and $T \leq 0.456$ respectively), we can identify the high and low shear segments in the transient and/or small strain region of the simulations. We then seek an effective temperature to describe the mesocluster size distributions observed in these segments separately. Where the system does \emph{not} undergo localisation, we use the mesocluster size distribution from the whole simulation box.

From Fig. \ref{figBandingStartAllTemps} we see that most of the systems with $T \geq 0.456$ exhibit localisation for a significant period of the simulation. Figure \ref{figBandingStart} shows that $T = 0.3$ typically does not do so until (approximately) the yield strain is reached or, if it does so, the period of localisation is very brief and we neglect these fluctuations. Thus, for $T \geq 0.456$, we analyse the high/low shear segments for the whole duration of the simulations, while for $T = 0.3$, we analyse the whole system at shear initialisation until localisation starts, after which point we analyse the high/low shear segments for the remaining duration of the simulation.

\begin{figure}
\begin{center}
\includegraphics[width=\columnwidth]{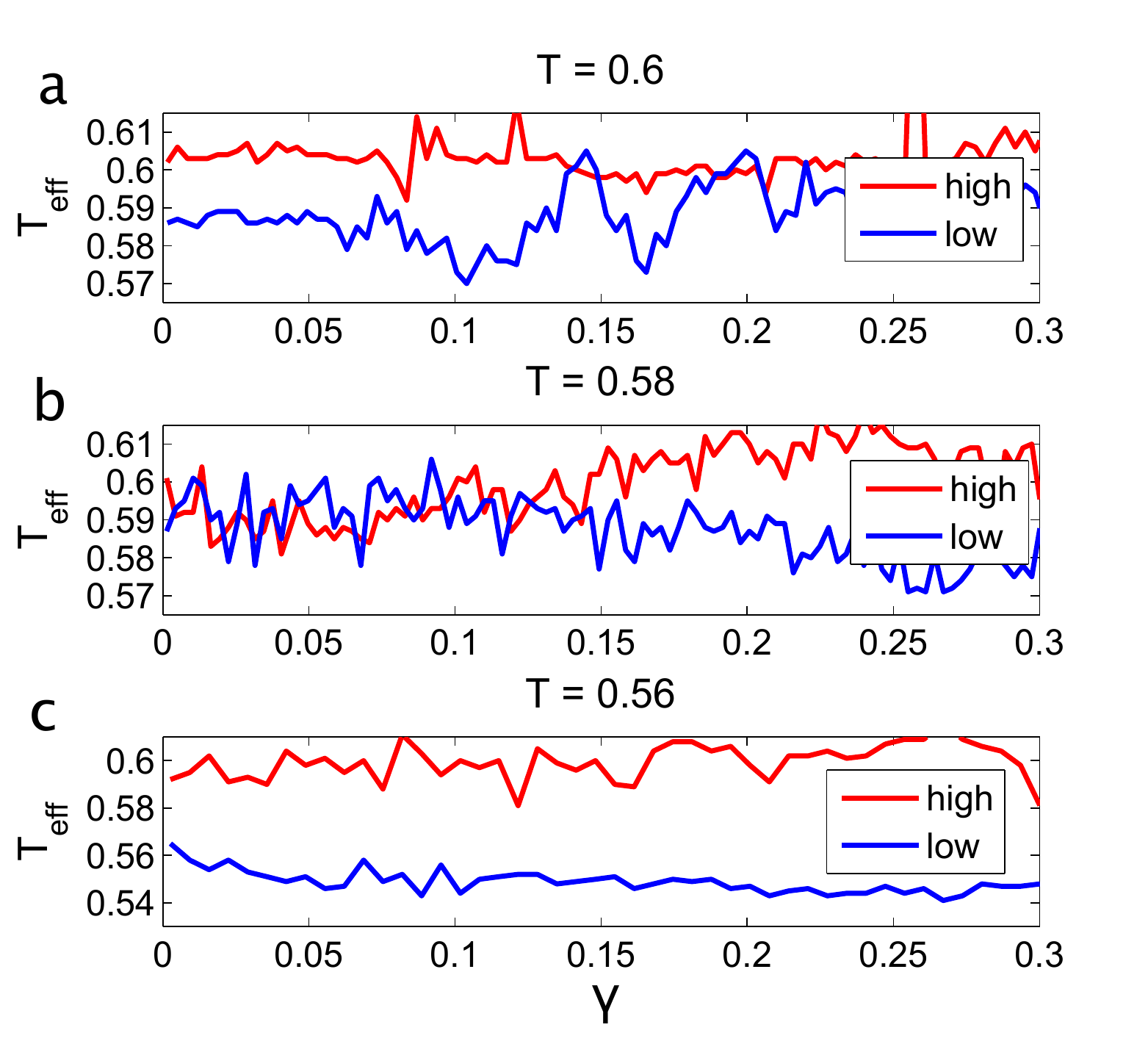}
\caption{The effective temperatures of the high and low shear segments for simulations with $T = 0.6, 0.58, 0.56$ and slowest shear rates studied ($10~{-5}$). Data are from single simulations corresponding to those used in Fig \ref{figTransientD2BigFig}. Overlaps correspond to regions where the bands are not necessarily clear and/or are too thin to accurately fit the mesocluster size distribution model.}
\label{figTransientTeff}
\end{center}
\end{figure}

Figure \ref{figTransientTeff} shows examples of the high and low shear segment effective temperatures (obtained using the mesocluster size distribution model) for the slowest shear rates available for $T = 0.6, 0.58, 0.56$. The low shear segments generally have lower effective temperatures than the high shear segments. Note that for $T = 0.58, 0.56$, the effective temperatures of the low shear band decrease as strain increases. For $T = 0.56$, the yield point occurs at $\gamma \approx 0.06$. We see that the effective temperature of the low shear band decreases during the transient regime ($\gamma \lesssim 0.06$) and stabilises as the steady state is approached at $\gamma \gg 0.06$.

\begin{figure}
\begin{center}
\includegraphics[width=\columnwidth]{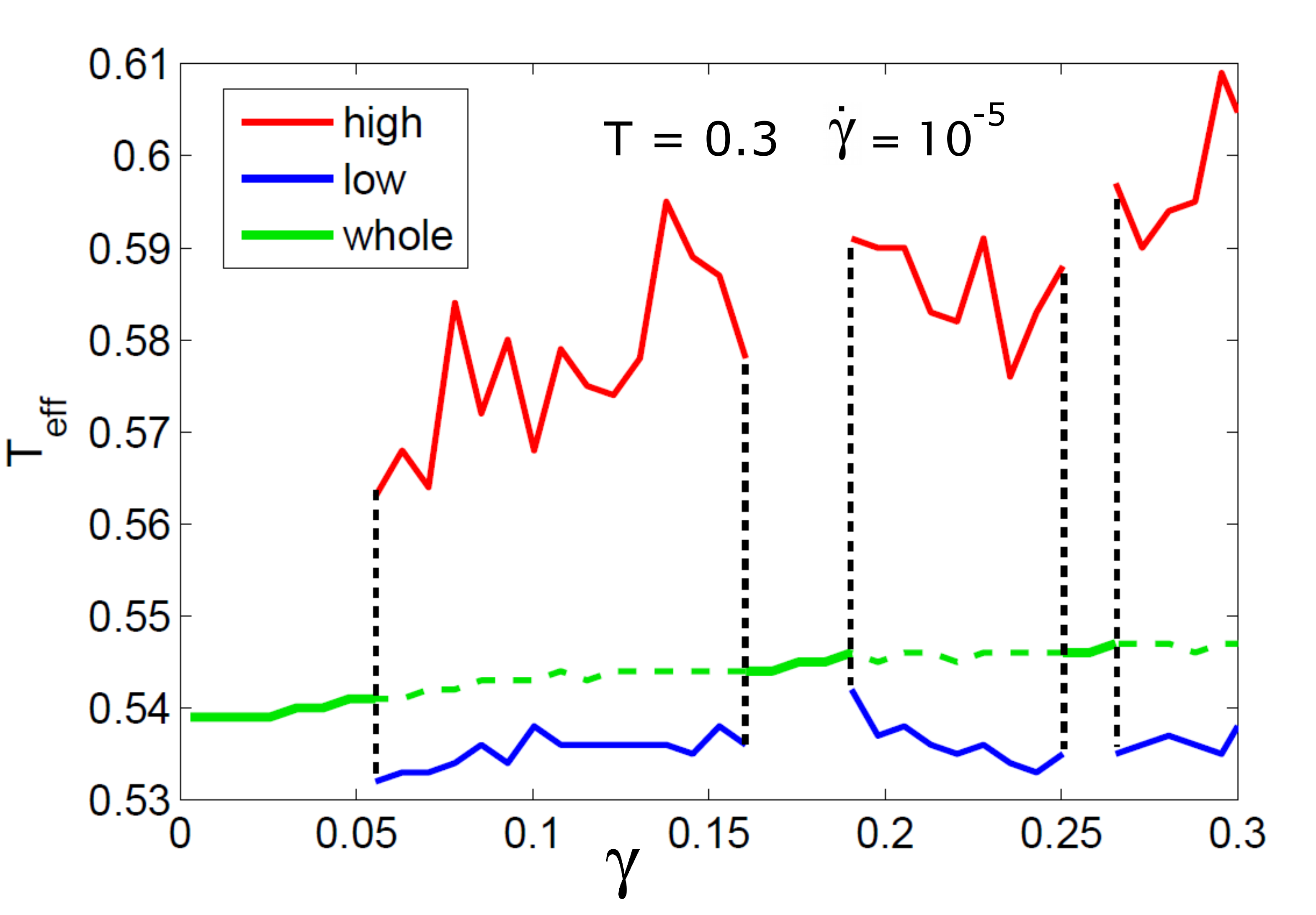}
\caption{The effective temperature of the whole system (green) and of the high (red) and low (blue) shear segments where the system exhibits localisation with $T = 0.3$ and shear rate $\dot{\gamma} = 10^{-5}$. Data from a single simulation of this state point. Dashed green lines track the effective temperature found from the mesocluster size distribution of the whole system. Clearly, the high and low segments follow the usual high/low temperature patterns, with the effective temperature of the high segments (and consequently the whole system) increasing steadily with increasing strain.}
\label{figT03TransientTeff}
\end{center}
\end{figure}

In Fig \ref{figT03TransientTeff}, we show the effective temperature of the whole system at state point $T = 0.3$, $\dot{\gamma} = 10^{-5}$, as it moves through the transient/small strain regime which then splits into two effective temperatures corresponding to the high and low shear segments close to yielding. Results from 
one initial configuration are shown, but this pattern is observed across all $T = 0.3$, $\dot{\gamma} = 10^{-5}$ configurations. The effective temperatures of the whole system and of the high shear segment increase steadily with increasing strain, while the low shear segment appears to stabilise as strain increases. The effective temperatures show no noticeable feature around the yield point. The only prominent feature close to yielding is the switch from a non-banded to a banded state.

\section{Discussion}
\label{sectionDiscussion}

We have studied the \emph{elastic} (and/or small strain) region of shear in the Wahnstr\"{o}m binary Lennard-Jones model over a wide temperature range for a variety of different shear rates. Our results are interpreted with a model based on the population of mesoclusters of icosahedra.

Like a number of other studies such as \cite{manning2007,candelier2009,puosi2016,rottler2014,nicolas2014},  we used the non-affine displacement parameter, $D^2$ ~ \cite{falk1998}, to define a criterion to find the regions of locally high/low shear rate. In all state points $T \geq 0.456$, the systems appear to start shear localisation instantaneously with the shear initialisation. However, for the state point $T = 0.3$, $\dot{\gamma} = 10^{-5}$, localisation¥ begins at strain values close to the yield point ($\approx 0.07$). The localisation signatures for this state point are consistent for different starting configurations. Despite the initial particle configurations being identical, different shear rates can result in significantly different shear localisation behavior. In particular, the small strain $D^2$ values for the slower shear rates for $T \geq 0.3$ are higher and less homogeneous than their faster shear counterpart. The slower shear rate of $\dot{\gamma} = 10^{-5}$ actually gave more shear localisation. This counterintuitive result has been obtained before in the case of slower quenches such as those we applied here \cite{shi2005}.

Using the $D^2$ values, we defined a threshold to determine whether or not a system was undergoing localisation. This threshold was defined on the typical range of $D^2$ values observed by segmenting the system along the $y$-direction. Sheared systems that exhibit larger ranges in the $D^2$ values than the quiescent systems are deemed to be undergoing shear localisation. The typical (normalised) range for quiescent systems were used as thresholds to define the beginning/end of shear localisation. Most state points studied exhibited localisation immediately, which largely persisted for the whole duration of the simulation of the transient/small strain regime and beyond yielding towards the steady state. Shear localisation was very strongly correlated with icosahedra poor regions (correlation coefficients generally between 0.65 and 0.85). The state point $T = 0.3, \dot{\gamma} = 10^{-5}$ started to band at strain values close to the yield point. The shear localisation was identified using the criterion from Ref. \cite{pinney2016}. As seen for all other state points, the local density of icosahedra correlated strongly with where the shear bands eventually formed. From these identified segments, an effective temperature was assigned to them according to the observed mesocluster size distributions. As expected for all systems \cite{manning2007,falk2011,pinney2016}, the high shear segments have higher effective temperatures than the low shear segments, with the difference between the segment effective temperatures being more pronounced at lower temperatures.

The values of the population of icosahedra, $\phi$ for our coldest temperature, $T = 0.3$, at large strain values increase slightly with decreasing shear rate, and all follow a similar pattern in the transient region; $\phi$ increases a small amount to reach a peak at a strain values of $\approx 0.01$, then falls at larger strain values. This is greatly exaggerated in the $T = 0.456$ data where for $\dot{\gamma} = 10^{-5}$, $\phi$ peaks at $\approx 0.05$ strain before decreasing at larger strain values, and peaks at an even higher strain value ($\approx 0.1$) when the shear rate is $\dot{\gamma} = 10^{-6}$. Results related to these have been observed previously, where the system can reach a deeper energy minimum with an imposed shear force \cite{lacks2004}. In such cases, the system can \emph{resemble} states lower in the energy landscape than the initial state.

Our results lead us to the following interpretation. Icosahedra are associated with local rigidity, and tend to resist shear. Thus the increase in population of icosahedra and development of extended mesoclusters at low temperature \cite{pinney2015} are consistent with the idea that LFS lead to rigidity in glassy systems. The observation that sheared systems can be related to quiescent systems at higher temperature, in that the mesocluster size distributions is very similar is remarkable in our opinion. We emphasise that determining the effective temperature is this way imposes a different level of constraint than does matching a single parameter. Furthermore in systems exhibiting shear localisation, each region, locally, has a different effective temperature, again with a well-described mesocluster size distribution.

Nevertheless, we caution that the response to shear is likely to be 
much more complicated than merely (somehow) generating configurations corresponding to different temperatures of the quiescent system. While our model does represent a reasonably complex description of the system (the size distribution of mesoclusters), it is far from complete. As noted above we neglect distortion and dilation of icosahedra \cite{albano2005}. More fundamentally, the fact remains that demonstrating a correlation between icosahedra and the mechanical properties of the material as we have does not in itself constitute a causal relation. Investigations in quiescent systems \cite{hocky2014,jack2014} suggest that other forms of ``amorphous order'' may likely be very important. One surprising finding, of immediate localisation upon slow shear (but not at higher shear) for very low temperatures remains unexplained. We leave this for future work, recalling that other forms of amorphous order and distortion/dilation of icosahedra should certainly be investigated.

A further important extension of our approach would be to investigate spatial correlation of the icosahedra and shear localisation. Our population dynamics model does include spatial correlation through the effects of a percolating network of icosahedra on the mesocluster size distribution \cite{pinney2015}. However we do not consider the effects of shear on such a network nor the spatial distribution of STZs \cite{nicolas2014,lin2016}. This is likely important, and would be a very interesting topic for future research. The combination of the kind of coarse-grained representation of local structure that we introduce here could in fact improve other coarse-grained models which may suffer form an underestimation of structural disorder \cite{nicolas2014}. Furthermore, our population dynamics mesocluster model has been explicitly developed to address temperatures far below those accessible to conventional numerical simulation \cite{pinney2015}. Thus it is possible to implement our model as a coarse-grained description of local structure, which could work for large systems at arbitrary temperature. Other important topics include the effect of distortion in icosahedra, \cite{albano2005} and the use of powerful simulation technique such as GPUs \cite{bailey2015} of particle swaps \cite{ninarello2017} to consider larger system sizes than has been possible here.

\section{Summary}
\label{sectionSummary}

In short, we have demonstrated that our mesocluster model approach to the transient shear regime may offer some predictive link between local structure (LFS) and the corresponding shear response. It was unexpected that yielding appeared to play a part in the localisation behavior of \emph{only one} state point, $T=0.3$.

We find that upon shear localisation, our mesocluster model may be fitted by assuming that the shear band is at a higher effective temperature than the rest of the material. Qualitatively similar behaviour has been seen in computer simulation \cite{bailey2006,nicolas2016}, while in experiment the situation remains unclear \cite{lewandowski2006,georgarakis2008,maass2015,thurnheer2016}. In the future it would be interesting to interrogate the prediction of the mesocluster model to see if the change in temperature predicted by the model corresponding to that observed in simulation \cite{bailey2006,nicolas2016}. 
However it is worth noting that in experiments \cite{slaughter2014}, while fracture on the microsecond timescale
does seem to correspond to significant temperature increases, on the milisecond shear localisation timescale, temperature changes are moderate at best.

\vspace{10mm}
\subsection*{Acknowledgements} 
The authors would like to thank Andrea Cavagna, Daniele Coslovich, Jens Eggers, Bob Evans, Rob Jack, Gilles Tarjus and Francesco Turci for many helpful discussions. We thank Peter Crowther for his artful rendering of an icosahedron. CPR would like to acknowledge the Royal Society for financial support and the European Research Council under the FP7 / ERC Grant agreement n$^\circ$ 617266 ``NANOPRS'', and Kyoto University SPIRITS fund. RP is funded by EPSRC grant code EP/E501214/1. This work was carried out using the computational facilities of the Advanced Computing Research Centre, University of Bristol.


\end{document}